\documentclass[preprint,showpacs,preprintnumbers,amsmath,amssymb]{revtex4}
\usepackage{epsf}
\usepackage{dcolumn}% Align table columns on decimal point
\usepackage{bm}% bold math
\everymath{\displaystyle}
\begin{document}

\title{Foldy-Wouthuysen transformation for relativistic
particles in external fields  }

\author{Alexander J. Silenko}

\affiliation{Institute of Nuclear Problems, Belarusian State
University,\\ Bobruiskaya str., 11, 220080 Minsk, Belarus}

\date{\today}

\begin {abstract}
A method of Foldy-Wouthuysen transformation for relativistic
spin-1/2 particles in external fields is proposed. It permits
determination of the Hamilton operator in the Foldy-Wouthuysen
representation with any accuracy. Interactions between a particle
having an anomalous magnetic moment and nonstationary
electromagnetic and electroweak fields are investigated.
\end{abstract}

 \maketitle

\section { INTRODUCTION}

   The Foldy-Wouthuysen (FW) representation \cite{FW} occupies special
place in the quantum theory. This is mainly due to the fact that
the Hamiltonian and all operators in this representation are
block-diagonal (diagonal in two spinors). For relativistic
particles in external fields, operators have the same form as in
the nonrelativistic quantum theory. Therefore, the FW
representation in the relativistic quantum theory is similar to
the nonrelativistic quantum theory. The basic advantages of the FW
representation are described in \cite{FW,CMcK} (see also below).

The transformation to the FW representation (FW transformation) holds only
in the one-particle approximation where the radiative corrections are not
calculated in a consistent way but are phenomenologically taken into account
by including extra terms in the Dirac equation (see \cite{F}).
One-particle description is feasible even for ultrarelativistic particles
if the external field is so weak that the probability of pair production or
bremsstrahlung losses can be neglected for a given interaction
energy of a particle. The range of applicability of this description is fairly
wide and includes, in
particular, the relativistic particle scattering and the
interaction of relativistic particles with matter and external fields.

In the nonrelativistic case, there exist a lot of good methods of FW
transformation with taking into account relativistic corrections \cite{FW,AB,E,KT}.
However, they are not useful for relativistic particles. The known methods of
solving this problem \cite{B,GS,STMP,Hol} either lead to cumbersome
calculations or the field
of their use is limited by the first approximation in field parameters. None of
these methods permits exact FW transformation for the particular cases
described in \cite{E,C,T}.
Therefore, the FW representation does not take the right stand in the
relativistic quantum theory. The Dirac and
Melosh \cite{Me} representations are mostly used.

FW transformation can also be performed for particles with spin $s>1/2$
\cite{C,Gue}.

In the present work, a method of FW transformation for relativistic particles
in external fields is proposed. This method permits obtaining a
Hamiltonian of any accuracy by successive approximations, as a power
series in the external field potentials and their derivatives. In some
cases, this method permits performing exact FW transformation.

   The relativistic system of units $\hbar=c=1$ is used.

\section { GENERAL PROPERTIES OF THE FOLDY-WOUTHUYSEN
REPRESENTATION}

   The basic advantages of the FW representation are due to its specific
properties.

   The relations between the operators in the FW representation are similar
to those between the respective classical quantities. In this representation,
the operators have the same form as in the nonrelativistic quantum theory.
Only the FW representation possesses these properties considerably simplifying
the transition to the semiclassical description. The FW
representation provides the best possibility of obtaining a meaningful classical limit of
the relativistic quantum theory.

For example, the Hamiltonian for a free particle fully agrees with that of
classical physics:
\begin{equation} {\cal H}_{FW}=\beta \sqrt{m^2+\bm p^2}, ~~~\bm p=-i\nabla,
\label{eq1} \end{equation}
in contrast with the Hamiltonian in the Dirac representation \cite{FW,NW}.
The position operator in the Dirac representation is the radius-vector,
$\bm r$ \cite{NW}. It corresponds to the mean position operator
for the free particle in the FW representation \cite{FW},
$$ \bm r_D=\bm r+\frac{i\beta\bm\alpha}{2\epsilon}-
\frac{i\beta(\bm\alpha\cdot\bm p)\bm p+[\bm\Sigma\times\bm p]p}
{2\epsilon(\epsilon+m)p},~~~p\equiv |\bm p|, ~~~\epsilon=\sqrt{m^2+p^2}. $$

   Here and below the following designations for the matrices are used:
$$\begin{array}{c}\bm{\gamma}=\left(\begin{array}{cc} 0  &  \bm{\sigma} \\ -\bm{\sigma} & 0
\end{array}\right), ~~~ {\beta}\equiv\gamma^0=\left(\begin{array}{cc} 1  &  0
\\ 0 & -1 \end{array}\right), ~~~\bm{\alpha}=\beta\bm\gamma=
\left(\begin{array}{cc} 0  &  \bm{\sigma} \\ \bm{\sigma} & 0
\end{array}\right), \\    \bm{\Sigma}
=\left(\begin{array}{cc} \bm{\sigma}  &  0 \\ 0 &
\bm{\sigma}\end{array}\right),   ~~~\bm{\Pi}=\beta\bm\Sigma
=\left(\begin{array}{cc} \bm{\sigma}  &  0 \\ 0 &
-\bm{\sigma}\end{array}\right),  \end{array}  $$
where $0,1,-1$ mean the corresponding 2$\times$2 matrices and $\bm{\sigma}$ is
the Pauli matrix.

In the FW representation, the problem of "{\em zitterbewegung}" motion never
arises \cite{CMcK,NW}. The operators $\bm l=\bm r\times\bm p$ and
$\bm\Sigma/2$ define the angular momentum and the spin for the free particle,
respectively.
In this representation, unlike the Dirac one, each of them is a constant of
motion (see \cite{FW}). The corresponding operators conserving in
the Dirac representation are
$$ \begin{array}{c}
\bm l_D=\bm r_D\times\bm p,\\
\frac{\bm\Sigma_D}{2}=\frac{\bm\Sigma}{2}-\frac{i\beta[\bm\alpha\times\bm p]}
{2\epsilon}-\frac{[\bm p\times[\bm\Sigma\times\bm p]]}
{2\epsilon(\epsilon+m)}.
\end{array}$$

The total angular momentum operator, $\bm j$, is a constant of motion
in both representations, because
$$ \bm j_D=\bm l_D+\frac{\bm\Sigma_D}{2}=\bm l+\frac{\bm\Sigma}{2}=\bm j. $$

The FW representation is very convenient for describing the particle polarization.
In this representation, polarization operators
have simple forms. For example, the three-dimensional polarization
operator equals the matrix $\bm\Pi$ \cite{FG,TKR}. In the Dirac representation,
this operator depends on the particle momentum \cite{FG,TKR}:
$$ \begin{array}{c}
\bm O\equiv \bm{\Pi_D}=\bm\Pi-\gamma^5\frac{\bm p}{\epsilon}-
\frac{\bm p(\bm\Pi\cdot\bm p)}{\epsilon(\epsilon+m)}.
\end{array} $$
For particles interacting with external fields, it also depends on the external
field parameters \cite{TKR}.

Thus, in the Dirac representation all operators corresponding to the basic
classical quantities are defined by cumbersome expressions.
These operators should also depend on the external field parameters
for particles interacting with external fields.

The FW representation helps one to prove that the particle position can
be measured up to its Compton wavelength \cite{FW,NW}. However, this property
is valid only for a particle not strongly interacting with external fields if
the one-particle approximation is attainable. Otherwise, the effect
of pair production prevents the use of both the Dirac equation (even with some
corrections) and the "traditional" Hamilton approach. Obviously, in this case
the FW transformation cannot be used either.

The FW transformation possesses another important property. The relativistic
wave equations and all operators are block-diagonal (diagonal in two
spinors). This property permits separating positive and negative energy states
\cite{FW}. Of course, extraction of even parts of operators becomes
unnecessary.

The detailed analysis performed in \cite{FPS} shows that the wave
functions in both the Dirac and FW representations are equal to each other only approximately,
and they do not coincide. In these papers, the nonrelativistic case was considered
and relativistic corrections was taken into account.

 An analogous conclusion follows from the results obtained in \cite{STMP}.
In this work, a more general situation has been investigated for the relativistic
particle not strongly interacting with an electromagnetic field. It has been found
that the upper spinors
in the Dirac and FW representations are approximately proportional to each
other, but this property is not exact.

Thus, the preferable employment of the FW representation is evident, although the
relativistic wave equations are more complicated in this representation.
\medskip\medskip

\section {METHODS OF THE FOLDY-WOUTHUYSEN TRANSFORMATION}

In the classical work by Foldy and Wouthuysen \cite{FW}, two different transformations,
for free relativistic particles and for nonrelativistic particles in electromagnetic
fields have been carried out. In the general case, transformation to a new
representation described by the wave
function $\Psi'$ is performed with the unitary operator $U$:
$$\Psi'=U\Psi=e^{iS}\Psi,$$
where $\Psi=\left(\begin{array}{c} \phi \\ \chi \end{array}\right)$ is the wave function (bispinor) in the Dirac
representation. As
 $$\Psi=U^{-1}\Psi',~~~i\frac{\partial}{\partial t}\Psi={\cal
H}\Psi, ~~~i\frac{\partial}{\partial t}\Psi'={\cal H}'\Psi', $$
the following transformation can be carried out:
$$\begin{array}{c} i\frac{\partial}{\partial
t}\Psi={\cal H}U^{-1}\Psi',~~~ i\frac{\partial}{\partial
t}\Psi=i\frac{\partial}{\partial t}\left(
 U^{-1}\Psi'\right)=i\frac{\partial U^{-1}}{\partial t}\Psi'+iU^{-1}
\frac{\partial\Psi'}{\partial t}\\= \left(i\frac{\partial
 U^{-1}}{\partial t}+U^{-1}{\cal H}'\right)\Psi',~~~U{\cal
 H}U^{-1}\Psi'=\left(iU\frac{ \partial U^{-1}}{\partial t}+{\cal
 H}'\right)\Psi'.\end{array}$$
   Hence, the Hamilton operator in the new representation
takes the form \cite{FW,Gol}:
\begin{equation} {\cal H}'=U{\cal H}U^{-1}-iU\frac{
\partial U^{-1}}{\partial t}, \label{eq2} \end{equation} or
$$ {\cal H}'=U\left({\cal H}-i\frac{\partial}{\partial
t}\right)U^{-1}+ i\frac{\partial}{\partial t}. $$

There is an error in this transformation in \cite{BD}.

   The Hamiltonian can be split into
operators commuting and noncommuting with the operator $\beta$:
\begin{equation} {\cal H}=\beta m+{\cal E}+{\cal
O},~~~\beta{\cal E}={\cal E}\beta, ~~~\beta{\cal O}=-{\cal O}\beta.
\label{eq3} \end{equation}

The Hamiltonian ${\cal H}$ is Hermitian. We assume that both operators
${\cal E}$ and ${\cal O}$ are also Hermitian.

For free Dirac particles ${\cal E}=0,~{\cal O}=\bm{\alpha}\cdot\bm{ p}$, and
the operator $S$ has the form
\begin{equation}  S=-i\beta\bm\alpha\cdot\bm p\theta(\bm{p}),
\label{eq4} \end{equation}
where $\theta$ is a function of the momentum operator. If we choose
$$ \theta(\bm{p})=\frac{1}{2p}\arctan{\biggl(\frac{p}{m}\biggr)},
$$
the transformed Hamiltonian ${\cal H}'$ contains no odd operators \cite{FW,BD}
and we obtain Eq. (1)
$$ {\cal H}'=\beta \sqrt{m^2+\bm p^2}. $$

   For nonrelativistic particles in an electromagnetic
field, the FW transformation can be performed with the operator \cite{FW,BD}
\begin{equation}  S=-\frac{i}{2m}\beta{\cal O}. \label{eq5} \end{equation}

The transformed Hamiltonian can be written in the form
\begin{equation}  \begin{array}{c}
{\cal H}'={\cal H}+i[S,{\cal H}]+\frac{i^2}{2!}[S,[S,{\cal
H}]]+\frac{i^3}{3!} [S,[S,[S,{\cal H}]]]+\dots\\-
\dot{S}-\frac{i}{2!}[S,\dot{S}]-\frac{i^2}{3!}
[S,[S,\dot{S}]]-\dots, \end{array} \label{eq6} \end{equation}
where $[\dots,\dots]$ means a commutator. As a result of this
transformation, we find
\begin{equation} {\cal H}'=\beta\epsilon+{\cal E}'+{\cal
O}',~~~\beta{\cal E}'={\cal E}'\beta, ~~~\beta{\cal O}'=-{\cal O}'\beta,
\label{eq7}\end{equation}
where the odd operator ${\cal O}'$ is now $ O(1/m)$. This procedure can be
repeated to obtain the required accuracy. Another form of the nonrelativistic
FW transformation was given by Ericksen \cite{E} (see also \cite{VJ}).

There are also other methods for obtaining the block-diagonal
form of the Hamiltonian or the Lagrangian. The so-called
elimination method of Pauli \cite{Pa} permits excluding the lower
spinor from relativistic wave equations. As a result, the wave
function of the final Pauli equation is the upper Dirac spinor,
$\phi$. This means that the upper Dirac spinor is also an eigenfunction
of the transformed Hamiltonian. However, this property is not
exact. The Pauli method was analyzed in detail in
\cite{FPS,VJ}. It was shown that this method gives the right first
approximation. Nevertheless, relativistic corrections of
higher orders are incorrect. It is quite natural because
direct Pauli's reduction leads to a neglect of the contribution of the
lower spinor \cite{FPS}. The relation between the exact wave
function in the FW representation and the upper Dirac spinor has
been found in \cite{STMP} in the relativistic case.

 A more exact variant of the elimination method had been proposed earlier by
Berestetskii and Landau \cite{BL} (see also \cite{AB,BLP}). They
showed that it was possible to find a nonunitary operator $V$ for which
\begin{equation} \psi=V\phi
\label{eq8} \end{equation}
is a two-component wave function with a correct norm. An appropriate form of
the operator $V$ can be obtained from the condition
$$ \int{\psi^\dag\psi dV}= \int{(\phi^\dag\phi+\chi^\dag\chi) dV}=1. $$
The relation between the Dirac spinors can be expressed in the general form:
$\chi=Q\phi$. Therefore,
$$V^\dag V=1+Q^\dag Q.$$

If we additionally assume that the operator $V$ is Hermitian, then both
this operator and the Hamiltonian can be found by successive approximations
\cite{AB,BL,BLP,St}.

Of course, the elimination method is much simpler. However, it is mostly intuitive.
Its validity is proved only by the coincidence of the results obtained by
the FW and Akhiezer-Berestetskii-Landau methods \cite{VJ}.

Another method of diagonalization of relativistic wave equations was
proposed by Korner and Thompson \cite{KT}. In this work, the Lagrangian
approach was used.
The Korner-Thompson method is similar to the FW method. It also includes a
successive decrease in the maximum order of odd terms. The results obtained by the FW
and Korner-Thompson methods agree (see \cite{Hol}).

Thus, several nonrelativistic transformation methods give the same results.
However, the FW transformation method has been justified in the best way.

In several cases, FW transformation can be performed exactly \cite{E,C,T}.
Exact FW transformation has also been performed for a wide class
of external fields in \cite{N}. In this work, involutive
symmetries of relativistic wave equations have been used. However,
the transformed Hamiltonians contain "nontraditional" space
reflection operators. The reduction of Hamiltonians to the
"traditional" form is a difficult problem. It has not been
investigated in \cite{N}. However, this reduction is
necessary to do for solving many problems (e.g., finding particle
and spin motion equations).

Generally, FW transformation for relativistic particles in external
fields is complicated. The transformation methods explained in \cite{B,GS}
require cumbersome calculations. A variant of the elimination method
useful for relativistic particles has been developed in \cite{STMP}. On
eliminating the lower spinor from the relativistic wave equations, the
final equation for the upper spinor takes the form \cite{STMP}:
\begin{equation} i\frac{\partial \phi}{\partial t}=F(\bm r,\bm p,
i\frac{\partial}{\partial t})\phi,
\label{eq9} \end{equation}
where $F$ is the operator function. Further calculations are analogous to those in
the Akhiezer-Berestetskii-Landau method. A new wave function with a
correct norm, $\psi$, expressed by Eq. (8) is introduced. Substituting it
for $\phi$ into
Eq. (9), one can find the Hamilton operator for the relativistic particle.

The relativistic wave equation for an upper spinor similar to Eq. (9) is
found by the Lagrangian approach \cite{Hol}.

However, it is difficult to find a second approximation by using the
relativistic variant of the elimination method proposed in \cite{STMP}.
It is easier to determine relativistic corrections of higher orders \cite{Hol}.

The right two-component wave function in the FW representation, $\psi$, does
not coincide with the upper Dirac spinor, $\phi$ \cite{STMP}. This conclusion is in
agreement with the results obtained in \cite{FPS}.

There are other difficult problems. The diagonalization of
relativistic wave equations needs carefulness, especially in the
time-dependent case. As mentioned in \cite{Gol,Ni}, in
the latter case ${\cal H}'$ is not equivalent to ${\cal H}$ since these
operators have different matrix elements. Rather, $U{\cal H}U^{-1}$ is.
There is a danger that one can arrive at a block-diagonal representation
differing from the FW one even in the time-independent
case. For example, the transformation performed in \cite{T}
(this is the Melosh transformation indeed \cite{W}) leads to a
block-diagonal Hamiltonian that differs from the Hamiltonian in
the FW representation \cite{YPhys}. Therefore, the application of
noncanonical transformation methods is restricted by the
necessity of verifying the results by comparing them with the
corresponding results obtained by the canonical transformation
method in some particular cases. Of course, other transformation
methods may be simpler or less cumbersome. Nevertheless, the FW
method is safer and substantiated very well.

In the present work, a relativistic extension to the FW method is proposed.

\section {EXACT FOLDY-WOUTHUYSEN TRANSFORMATION}

   Consider some cases of the exact FW transformation.

In Eq. (3), the operators $\beta$ and ${\cal O}^2$ commute ($\beta{\cal O}^2=-{\cal O}\beta
{\cal O}={\cal O}^2\beta$). Therefore, the operator
${\cal O}^2$ is even.

The operator $S$ can be defined by an expression similar to Eq. (4):
\begin{equation}
S=-i\frac{\beta{\cal O}}{C}\theta, \label{eq10}
\end{equation}
where $C$ and $\theta$ are the functions of ${\cal O}^2$ and the operator $C$ satisfies
the following conditions:
\begin{equation}
C^2={\cal O}^2, ~~~~~~~[\beta,C]=0.
\label{eq11}
\end{equation}
It follows from conditions (11) that the operator $C$ is also even.

It is possible to use the following formal definition of this operator:
\begin{equation}
C=\sqrt{{\cal O}^2}.
\label{eq12}
\end{equation}

Relations (11),(12) define the
square root of matrix operators. To unambiguously define the square root,
these relations should be complemented by the condition that the
square root of the unit matrix ${\cal I}$ is equal to the unit matrix.
This definition of the square root coincides with those of
\cite{FW,BD,N}.
For example, for free particles
$$ {\cal O}=\bm\alpha\cdot\bm p,~~~ {\cal O}^2={\cal I}\bm p^2,
~~~C={\cal I}\sqrt{\bm p^2}\equiv {\cal I}|\bm p|. $$
Further, the symbol of the unit matrix ${\cal I}$ will be omitted.

Since $${\cal O}^2=-\beta{\cal O}\beta{\cal O},~~~C=\sqrt{-\beta{\cal
O}\beta{\cal O}},~~~f({\cal O}^2)=f(C^2),$$
the operators $\beta{\cal O},C,{\cal O}^2$, and $\theta$ commute with
each other.
The operator $\theta$ is the angle of rotation of the basic vector set in the
spinor space.

   The unitary transformation operator, $U$, can be written in the form
\begin{equation}
U=\cos \theta+\frac{\beta{\cal O}}{C}\sin \theta.
\label{eq13}
\end{equation}

An FW transformation is exact if the external
field is stationary and the operators ${\cal E}$ and ${\cal O}$
commute:
\begin{equation}
[{\cal E},{\cal O}]=0.
\label{eq14}
\end{equation}

In this particular case,
$$ [{\cal E},\beta{\cal O}]=\beta[{\cal E},{\cal O}]=0. $$

Condition (14) is a sufficient but not necessary condition of the
exact transformation.

The Hamilton operator in the new representation takes the form
$$\begin{array}{c}
{\cal H}'=\left(\cos\theta+\frac{\beta{\cal O}}{C}\sin
\theta\right) {\cal H}\left(\cos\theta-\frac{\beta{\cal O}}{C}\sin
\theta\right)\\ =(\beta m+{\cal
O})\left(\cos\theta-\frac{\beta{\cal O}}{C}\sin \theta\right)^2+
{\cal E} =(\beta m+{\cal O})\left(\cos 2\theta-\frac{\beta{\cal
O}}{C} \sin 2\theta\right)+{\cal E}\\= \beta\left(m\cos
2\theta+C\sin 2\theta\right)+{\cal O}\left(\cos 2\theta-\frac mC
\sin 2\theta\right)+{\cal E}.\end{array}$$

The Hamiltonian ${\cal H}'$ is even if the odd term (proportional to
${\cal O}$) vanishes. This takes place if
\begin{equation}
\tan 2\theta=\frac Cm.
\label{eq15}
\end{equation}

This equation has two solutions, $\theta_1$ and $\theta_2$, differing in $\pi/2$.
Since
$$\tan 2\theta=\frac{2\tan\theta}{1-\tan^2\theta},~~~\tan\theta=
\frac{\tan 2\theta}{1\pm\sqrt{1+\tan^2 2\theta}},   $$
they are defined by the relations
\begin{equation}
\tan \theta_1=\frac {C}{\epsilon+m},~~~
\tan \theta_2=-\frac {C}{\epsilon-m},~~~\epsilon=\sqrt{m^2+C^2}
=\sqrt{m^2+{\cal O}^2}.
\label{eq16}
\end{equation}

Thus, there are two unitary transformations of the operator
${\cal H}$ to an even form. They are characterized by the
angles $\theta_1$ and $\theta_2$, where the angle $\theta_1$ corresponds to the
FW transformation.

   As a result of both transformations, one of the spinors (lower for
$\theta_1$ and upper for $\theta_2$) becomes zero as for free particles.

   Note that the transformation under consideration is also similar
to the Melosh transformation \cite{Me}.

 If condition (14) is satisfied, then the
Hamilton operator in the FW representation is defined exactly:
\begin{equation}
{\cal H}'=\beta \epsilon+{\cal E}.
\label{eq17}
\end{equation}

Unlike \cite{B,GS,STMP}, Eq. (17) contains exact expressions
for the Hamiltonian derived in \cite{E,C,T} as particular cases.

   The transformation operator $U$ can be written in nonexponential form.
After the calculation of $\sin\theta_1$ and $\cos\theta_1$ with formulae (16),
$$\sin\theta_1=\frac{C}{\sqrt{2\epsilon(\epsilon+m)}},~~~
\cos\theta_1=\sqrt{\frac{\epsilon+m}{2\epsilon}},   $$
we obtain the following expression:
\begin{equation}
U^{\pm}=\frac{\epsilon+m\pm\beta{\cal O}}{\sqrt{2\epsilon(\epsilon+m)}}.
\label{eq18} \end{equation}
where $U^+\equiv U,~U^-\equiv U^{-1}$. This expression
agrees with the well-known formula for free particles \cite{FW}. Since
$(\beta{\cal O})^{\dag}={\cal O}\beta=-\beta{\cal O}$, the operator $U$ is
unitary. Simultaneous change of signs of $\sin\theta_1$ and $\cos\theta_1$
does not affect the final result because the wave functions $\Psi,\Psi'$
are determined up to a sign.
Direct calculation of the Hamilton operator in the FW representation
also leads to Eq. (17) in accordance with formulae (2),(3),(14),(18).

Another class of Hamiltonians permitting
exact FW transformation has been investigated in \cite{N}.

\section {EXACT TRANSFORMATION FOR PARTICLES IN ELECTROWEAK\\
FIELDS}

   Let us consider the interaction of a relativistic
spin-1/2 particle, possessing an anomalous magnetic moment (AMM), with
stationary electromagnetic and electroweak fields.
The Hamiltonian of the electromagnetic interaction is defined by the
Dirac-Pauli equation \cite{P}:
\begin{equation} {\cal H}_{DP}=\bm{\alpha}\cdot\bm{\pi}+\beta
m+e\Phi+\mu'(-\bm {\Pi}\cdot \bm{H}+i\bm{\gamma}\cdot\bm{
E}),~~~\bm{\pi}=\bm{p}-e\bm{ A},
\label{eq19}
\end{equation}
where $\mu'$ is AMM, $\Phi,\bm{ A}$ and $\bm{ E},\bm{H}$ are the
potentials and the strengths of an electromagnetic field.
This equation is derived in the one-particle
approximation and is useful when an electromagnetic
field is not extremely strong (see \cite{STMP}).

   The weak interaction Hamiltonian should be added to the Hamiltonian
of Eq. (19). The weak interaction does not conserve the spatial parity.
For the interaction transferred by neutral currents, the standard
model gives the following expression for the
parity-nonconserving weak interaction Hamiltonian in the
approximation of a small transferred momentum \cite{CB}:
\begin{equation}
{\cal H}_{PNC}=-\frac{G}{\sqrt2}\left( C_1\gamma^5+C_2\bm{\alpha}\cdot\bm{
\sigma}'\right)n(\bm{ r}),~~~\gamma^5=\left( \begin{array}{cc} 0  & -1
\\ -1  & 0 \end{array}\right), \label{eq20} \end{equation}
where $G$ is
the Fermi constant, $\bm{\sigma}'$ is the Pauli matrix for matter
particles, and $n(\bm{ r})$ is the density of matter particles. For the
interactions with nuclei, $n(\bm{ r})$ characterizes the density
of nucleons of a certain kind, and $\bm{\sigma}'$ should be replaced by
the nucleus spin. Formulae (19),(20) do not change if the external fields are
nonstationary. The matter particles are considered to be at rest.

The coefficients $C_1,C_2$ are different for different
pairs of interacting particles. The Hamiltonians corresponding to the
interactions with different matter particles should be summed-up. The
signs in formula (20) depend on the definition of the coefficients
$C_1,C_2$ and matrix $\gamma^5$. The total Hamiltonian
equals
\begin{equation}{\cal H}= {\cal H}_{DP}+{\cal H}_{PNC}.
\label{eq21}\end{equation}
In this case, in formulae (3),(16)--(18) we have
\begin{equation}\begin{array}{c} {\cal E}=e\Phi-\mu'\bm {\Pi}\cdot\bm{H},
~~~~~~~ {\cal O}=\bm{\alpha}\cdot\bm{\pi}+
i\mu'\bm{\gamma}\cdot\bm{E}-\frac{G}{\sqrt2}\left( C_1\gamma^5+C_2
\bm{\alpha}\cdot\bm{\sigma}'\right)n(\bm{ r}) \\
=\beta\left[\bm{\gamma}\cdot \bm{\pi}+i\mu'\bm{\alpha}\cdot\bm{
E}-\frac{G}{\sqrt2}\left( C_1\beta
\gamma^5+C_2\bm{\gamma}\cdot\bm{\sigma}'\right)n(\bm{ r})\right].
\end{array}\label{eq22}\end{equation}

Let us consider some particular cases where Hamiltonian (21) satisfies
condition (14). For these cases the FW transformation
is exact. The general case will be analyzed in the next section.

The exact Hamiltonian in the FW
representation is given by Eq. (17), where ${\cal E}$ is defined
by Eq. (22), and
\begin{equation} \begin{array}{c}
\epsilon=\Biggl\{m^2-\left[
\bm{\gamma}\cdot\bm{\pi}+i\mu'\bm{\alpha}\cdot\bm{
E}-\frac{G}{\sqrt2} \left(
C_1\beta\gamma^5+C_2\bm{\gamma}\cdot\bm{\sigma}'\right)n(\bm{ r})
\right]\Biggr\}^{1/2}\\= \Biggl\{m^2+\bm{\pi}^2+\beta\mu'
\left(\bm\Sigma\cdot[\bm\pi\!\times\!\bm E]-\bm\Sigma\cdot[\bm
E\!\times\! \bm\pi]-\nabla\!\cdot\!\bm E\right)+\mu'^2\bm
E^2-e\bm\Sigma\cdot\bm H
\\ \left.+
\frac{G}{\sqrt2}\biggl(C_1\{\bm{\Sigma}\!\cdot\!\bm {\pi},n(\bm{
r})\}_+- C_2\{\bm{\sigma}'\!\cdot\!\bm{\pi},n(\bm{r})\}_+
+C_2[\bm{\Sigma}\!\times\!\bm{\sigma}']\!\cdot\!\nabla n(\bm{
r})\right.\\ -
2\beta\mu'C_2[\bm{\Sigma}\!\times\!\bm{\sigma}']\!\cdot\!\bm E
n(\bm{ r})\biggr)+\frac{G^2}{2}n^2(\bm{ r})\left[C_1^2+
3C_2^2-2C_2(C_1+C_2)\bm{\Sigma}\cdot\bm{\sigma}'\right]\Biggr\}^{1/2}.
\end{array}
\label{eq23}
\end{equation}

Hence, the operator $U^{\pm}$ is expressed by formulae
(18),(22),(23). Although Eq. (23) is formally exact, the small terms
proportional to $C_1^2,C_1C_2,C_2^2$ are wittingly negligible
in the approximation of a small transferred momentum.

   Formulae (17),(22),(23) describe the exact Hamilton operator
in the FW representation in the following particular cases:

~~~a) in the presence of only weak interaction
($\Phi\!=\!0, \bm A\!=\!0, \bm E\!=\!0, \bm H\!=\!0$);

~~~b) for Dirac particles ($\mu'=0$) in magnetic and weak fields
($\Phi\!=\!0,\bm E\!=\!0$);

~~~c) for uncharged particles with AMM in electric and weak
fields ($e\!=\!0,\bm A\!=\!0,\bm H\!=\!0$);

~~~d) for particles with AMM moving in the plane orthogonal
to a static uniform magnetic field ($\Phi\!=\! 0,\bm E\!=\!0,
P_z \! =\! 0, C_1\!=\!C_2\!=\!0$);

~~~e) for uncharged particles with AMM moving in the plane
orthogonal to a static uniform magnetic field. A static
electric field (possibly nonuniform) is also orthogonal
to the magnetic field ($e\!=\! 0,\bm E\!\perp\!\bm
H,P_z\!=\! 0, C_1\!=\!C_2\!=\!0$).

In two cases (d) and e)), $\bm H\!=\!H\bm e_z$, and in the case e), the electric
field strength does not depend on
$z$. Otherwise, ${\rm rot}\bm E\!\neq\! 0$
and the magnetic field is not constant ($\partial\bm
H/\partial t\!\neq\! 0$). Therefore, in these cases the operator
$p_z\!=\!-i(\partial /\partial z)$ commutes with the Hamilton operator
and has eigenvalues
$P_z\!=\!{\rm const}$. Consequently, the consideration of the particular case
$P_z\!=\!0$ is quite reasonable. All these cases satisfy condition (14).

Formulae (17),(22),(23)
agree with all exact expressions of the operator ${\cal H}'$ derived
for uncharged particles with AMM in an electrostatic field,
Dirac particles in a static magnetic field,
and particles with AMM moving in the plane
orthogonal to a static uniform magnetic field in
\cite{E,C,T}. The weak interaction is not considered in these works.

\section {GENERAL CASE}

   In the general case, relativistic particles interact with external fields.
We suggest to perform the FW transformation in two stages.
First, a transformation similar to the FW transformation for free particles,
is performed for particles in external fields. Second, a transformation
similar to the FW transformation for nonrelativistic particles is carried out.

   We assume that the external fields are not extremely strong and the
transformed Hamiltonian can be expressed as a power series in the field
potentials and their derivatives. The external fields can be nonstationary.

   In the general case, formula (17) is not exact because ${\cal E}$ depends
on the coordinates and contains Dirac matrices. We
should calculate the commutator of the operators $U$ and ${\cal
E}\! -\! i(\partial/\partial t)$:
$$U\biggl({\cal E}-i\frac{\partial}{\partial t}\biggr) U^{-1}=
{\cal E}-i\frac{\partial}{\partial t}+\biggl[U,{\cal E}-
i\frac{\partial}{\partial t}\biggr]U^{-1}.$$

   In this case, it is necessary to compute some commutators containing inverse
operators and square roots of the operators. These
   commutators can be calculated using the following exact formulae which are
valid for arbitrary operators $A$ and $B$ \cite{SJETP}:
\begin{equation} [A^{-1},B]=A^{-1}[B,A]A^{-1},
\label{eq24}\end{equation} \begin{equation}
[A,B]=\frac14\{A^{-1},[A^{2},B]\}_+-\frac14\left[[A,
[A,B]],A^{-1}\right],
\label{eq25}\end{equation}
where $A^{-1}\equiv 1/A$ and $\{\dots,\dots\}_+$ stands for the anticommutator.
If $A$ is the square root of the operators and the commutator of the operators is
small compared to their product, i.e.,
$$ |[A,B]| \ll |AB|, $$
formulae (24),(25) allow us to obtain the quantity $[A,B]$ with
any accuracy by the method of successive approximations (see
\cite{SJETP}). As a rule, this condition is satisfied since it is
equivalent to the inequality
\begin{equation}
\frac{\hbar c}{E}\ll l_c,
\label{eq26}\end{equation}
where$E$ is the total energy including the rest energy and $l_c$ is
the characteristic size of the nonuniformity region of the external field. For
the nonrelativistic particle, the quantity $\hbar c/E$ is equal to the Compton
wavelength.

First, it is necessary to perform a unitary transformation with operator (18).
After this operation, the
Hamiltonian ${\cal H}'$ still contains odd terms proportional to
the derivatives of the potentials. Let us write the operator ${\cal H}'$ as
\begin{equation} {\cal H}'=\beta\epsilon+{\cal E}'+{\cal
O}',~~~\beta{\cal E}'={\cal E}'\beta, ~~~\beta{\cal O}'=-{\cal O}'\beta,
\label{eq27}\end{equation}
where
\begin{equation}  \begin{array}{c}
\epsilon=\sqrt{m^2+{\cal O}^2}, \\
{\cal E}'=i\frac{\partial}{\partial t}+\frac{\epsilon+m}
{\sqrt{2\epsilon(\epsilon+m)}}\left({\cal E}-i\frac{\partial}{\partial t}
\right)\frac{\epsilon+m}{\sqrt{2\epsilon(\epsilon+m)}}-\frac{\beta{\cal O}}
{\sqrt{2\epsilon(\epsilon+m)}}\left({\cal E}-i\frac{\partial}{\partial t}
\right)\frac{\beta{\cal O}}{\sqrt{2\epsilon(\epsilon+m)}}, \\
{\cal O}'=\frac{\beta{\cal O}}{\sqrt{2\epsilon(\epsilon+m)}}
\left({\cal E}-i\frac{\partial}{\partial t}
\right)\frac{\epsilon+m}{\sqrt{2\epsilon(\epsilon+m)}}-
\frac{\epsilon+m}{\sqrt{2\epsilon(\epsilon+m)}}\left({\cal E}-i\frac{\partial}
{\partial t}\right)\frac{\beta{\cal O}}{\sqrt{2\epsilon(\epsilon+m)}}.
\end{array} \label{eq28} \end{equation}

Since
$$ ABA=\frac12\left(\{A^{2},B\}_+-[A,[A,B]]\right), $$
relation (28) for the operator ${\cal E}'$ takes the form
\begin{equation}  \begin{array}{c}
{\cal E}'={\cal E}-\frac14\left[\frac{\epsilon+m}
{\sqrt{2\epsilon(\epsilon+m)}},\left[\frac{\epsilon+m}
{\sqrt{2\epsilon(\epsilon+m)}},\left({\cal E}-i\frac{\partial}{\partial t}
\right)\right]\right] \\
+\frac14\left[\frac{\beta{\cal O}}
{\sqrt{2\epsilon(\epsilon+m)}},\left[\frac{\beta{\cal O}}
{\sqrt{2\epsilon(\epsilon+m)}},\left({\cal
E}-i\frac{\partial}{\partial t} \right)\right]\right].
\end{array} \label{eq29}\end{equation}

The odd terms are small compared to both $\epsilon$ and the initial
Hamiltonian ${\cal H}$. This circumstance
allows us to apply the usual scheme of the nonrelativistic
FW transformation \cite{FW,BD}.

Second, the transformation should be performed with the following operator:
\begin{equation} U'=\exp{(iS')}, ~~~
S'=-\frac i4\beta\left\{{\cal
O}',\frac{1}{\epsilon}\right\}_+=-\frac i4\left[\frac{\beta}{\epsilon},
   {\cal O}'\right]. \label{eq30} \end{equation}

   The further calculations are similar to those given in
\cite{BD}. The particle mass should be replaced by the operator
$\epsilon$ noncommuting with the operators
${\cal E}',{\cal O}'$. If only major corrections are taken into account, then
the transformed Hamiltonian equals
\begin{equation} {\cal H}''=\beta\epsilon+{\cal
   E}'+\frac14\beta\left\{{\cal O}'^2,\frac{1}{\epsilon}\right\}_+.
\label{eq31} \end{equation}

This is the Hamiltonian in the FW representation.

  To obtain the desired accuracy, the calculation
procedure with the transformation operator (30) ($S'$ is replaced by
$S'',S'''$ {\em etc}.) should be repeated multiply.

     Let us calculate the Hamiltonian in the FW representation for the
relativistic particle with AMM interacting with a nonstationary
   electroweak field. The Hamiltonian in the Dirac representation
   is defined by formulae (19)--(21). The transformed
Hamiltonian is defined by
Eq. (30), where the operator ${\cal O}'$ contains the field
strengths and does not contain the field potentials. Let us deduce the Hamiltonian
to within first-order terms in the field strengths and their
first derivatives and second-order terms in the field potentials.
The terms of the second order and higher in the field strengths and their
derivatives and the first-order terms containing derivatives of the second order
and higher of the field strengths will be omitted.

   Since we neglect the second-order quantities in ${\cal O}'$, the
operator ${\cal O}'$ does not make any contribution to the Hamiltonian
${\cal H}''$ at the second stage of transformation defined by formula
(31). As a result, we obtain the following equation for the Hamiltonian
in the FW representation:
   \begin{equation}\begin{array}{c} {\cal H}''=\beta\epsilon+{\cal E}',\\
{\cal E}'=e\Phi+\frac
   e8\left\{\frac{1} {\epsilon(\epsilon
   +m)},\left(\bm\Sigma\!\cdot\![\bm\pi\!\times\!\bm E]-\bm\Sigma
\!\cdot\![\bm E\!\times\!\bm\pi]-\nabla\!\cdot\!\bm
E\right)\right\}_+ \\
+\frac{e}{32}\left\{\frac{2\epsilon^2+2\epsilon m+m^2}{\epsilon^4(
\epsilon +m)^2},\bm\pi\!\cdot\!\nabla(\bm\pi\!\cdot\!\bm E+\bm
E\!\cdot\!\bm\pi)\right\}_+-\mu' \bm\Pi\!\cdot\!\bm H\\
+\beta\frac{\mu'}{4}\left\{\frac{1}{\epsilon(\epsilon+m)},
\biggl[(\bm{H}\!\cdot\!\bm\pi)(\bm{\Sigma}\!\cdot\!\bm\pi)+
(\bm{\Sigma}
\!\cdot\!\bm\pi)(\bm\pi\!\cdot\!\bm{H})+2\pi(\bm\pi\!\cdot\!\bm j+
\bm j\!\cdot\! \bm\pi)\biggr]\right\}_+,
\end{array} \label{eq32} \end{equation}
where $\bm j=\nabla\!\times\!\bm H-\frac{1}{4\pi}\frac{\partial \bm E}
{\partial t}$
is the external current density, and $\epsilon$ is determined by formulae (23),(27).
It is important that the operators $\epsilon,{\cal E}'$ are found at
the first stage, i.e., at the transformation with operator (18).

   In the weak field approximation,
$$ \begin{array}{c}
\epsilon=\epsilon'+\beta\frac{\mu'}{4}\left\{\frac{1}{\epsilon'},\biggl(
\bm\Sigma\cdot[\bm\pi\!\times\!\bm E]-\bm\Sigma\cdot[\bm
 E\!\times\!\bm\pi]-\nabla\!\cdot\!\bm E\biggr)\right\}_+\\
-\frac{e}{4}\left\{\frac{1}{\epsilon'},\bm\Sigma\cdot\bm H\right\}_++
 \frac{G}{4\sqrt2}\left\{\frac{1}{\epsilon'},W\right\}_+
\end{array} $$
and
\begin{equation}\begin{array}{c} {\cal H}''=\beta\epsilon'+e\Phi+\frac
   14\left\{\left(\frac{\mu_0m}{\epsilon'
   +m}+\mu'\right)\frac{1}{\epsilon'},\biggl(\bm\Sigma\!\cdot\![\bm\pi\!
\times\!\bm E]-\bm\Sigma\!\cdot\![\bm E\!\times\!\bm\pi]-\nabla\!
\cdot\!\bm E\biggr)\right\}_+\\ +\frac{\mu_0m}{16}
\left\{\frac{2\epsilon'^2+2\epsilon' m+m^2}{\epsilon'^4( \epsilon'
+m)^2},\bm\pi\!\cdot\!\nabla(\bm\pi\!\cdot\!\bm E+\bm
E\!\cdot\!\bm\pi)\right\}_+-\frac
12\left\{\left(\frac{\mu_0m}{\epsilon'}
+\mu'\right), \bm\Pi\!\cdot\!\bm H\right\}_+\\
+\frac{\mu'}{4}\left\{\frac{1}{\epsilon'(\epsilon'+m)},
\biggl[(\bm{H}\!\cdot\!\bm\pi)(\bm{\Pi}\!\cdot\!\bm\pi)+ (\bm{\Pi}
\!\cdot\!\bm\pi)(\bm\pi\!\cdot\!\bm{H})+2\pi(\bm\pi\!\cdot\!\bm j+
\bm j\!\cdot\! \bm\pi)\biggr]\right\}_+
+\frac{G}{4\sqrt2}\left\{\frac{1}{\epsilon'},W\right\}_+,
\end{array} \label{eq33} \end{equation}
where
\begin{equation}
\epsilon'=\sqrt{m^2+\bm{\pi}^2},~~~ W=C_1\{\bm{\Sigma}\cdot\bm {\pi},
n(\bm{ r})\}_+-C_2\{\bm{\sigma}'\cdot\bm{\pi},n(\bm{r})\}_+
+C_2[\bm{\Sigma}\!\times\!\bm{\sigma}']\!\cdot\!\nabla n(\bm{ r}),
\label{eq34} \end{equation}
and $\mu_0=e/(2m)$ is the Dirac magnetic moment.

   Unlike works \cite{B,GS,STMP}, formula (33) includes, as particular
cases, the exact expressions for the Hamiltonian in the FW
representation obtained in \cite{E,C,T}. Formulae (32)--(34)
also agree with the results obtained in \cite{B,GS,STMP,L,STM}.
Detailed analysis shows that the method of FW transformation used
in \cite{STMP} does not allow one to take into consideration the terms
proportional to the double commutators of $\epsilon$ with
$e\Phi$ and $\mu'\bm\Pi\!\cdot\!\bm H~\bigl(e[\epsilon,[\epsilon,\Phi]]$
and $\mu'[\epsilon,[\epsilon,\bm\Pi\!\cdot\!\bm H]]\bigr)$. The terms
proportional to $\bm E$ and $\bm H$ in the Hamiltonian obtained in
\cite{STMP} coincide with those in \cite{B,GS} and Eq. (33). However,
only the Dirac particles were considered in \cite{B}, the derivatives of the field
strengths were neglected in \cite{GS}, and the nonrelativistic Hamiltonian
with relativistic corrections was found in \cite{L}.

\section {PARTICLE AND SPIN MOTION EQUATIONS}

Thus, the FW representation is very convenient for describing
the particle and spin motion owing to the simple forms of operators. In
order to derive corresponding quantum equations, it is
necessary to compute the commutators of the Hamiltonian with the
same operators as in the nonrelativistic theory. The kinetic
momentum operator of particles in an electromagnetic field equals
$\bm\pi=\bm p-e\bm A$ \cite{foo}. The equation of particle motion
in the electromagnetic field is defined in terms of the commutator of the
Hamiltonian with this operator:
$$\frac{d\bm\pi}{dt}=i[{\cal H}'',\bm\pi]-e\frac{\partial\bm A}{dt}. $$

To determine the quantum equation of particle motion, we take into account
Eq. (33) for the Hamiltonian
\begin{equation}   \begin{array}{c}
\frac{d\bm \pi}{dt}=e\bm E+\beta\frac{e}{4}\Biggl\{\frac{1}{\epsilon'},
\biggl([\bm\pi\times\bm H]-[\bm H\times\bm\pi]\biggr)\Biggr\}_+\\
+\frac 14\Biggl\{\Biggl(\frac{\mu_0m}{\epsilon'+m}+\mu'\Biggr)
\frac{1}{\epsilon'},\Biggl[\nabla(\bm\Sigma\cdot[\bm
E\times\bm\pi])-
\nabla(\bm\Sigma\cdot[\bm\pi\times\bm E])+\Delta\bm E\Biggr]\Biggr\}_+ \\
-\frac{\mu_0m}{16}\Biggl\{\frac{2\epsilon'^2+2\epsilon' m+m^2}
{\epsilon'^4(\epsilon'+m)^2},(\bm\pi\cdot\nabla)\nabla(\bm
\pi\cdot\bm E+ \bm E\cdot\bm \pi)\Biggr\}_++
\frac{1}{2}\Biggl\{\Biggl(\frac{\mu_0m}{\epsilon'}+\mu'\Biggr),
\nabla(\bm\Pi\cdot\bm H)\Biggr\}_+\\
-\frac{\mu'}{4}\Biggl\{\frac{1}{\epsilon'(\epsilon'+m)},\Biggl[(\bm{\Pi}
\!\cdot\!\bm\pi)\nabla(\bm\pi\!\cdot\!\bm{H})+\biggl(\nabla(\bm{H}
\!\cdot\!\bm\pi)\biggr)(\bm\Pi\!\cdot\!\bm\pi)+
2\pi\nabla(\bm\pi\!\cdot\!\bm j+\bm j\!\cdot\!
\bm\pi)\Biggr]\Biggr\}_+.
\end{array}  \label{eq35} \end{equation}

The equation of spin motion is defined by the formula
$$\frac{d\bm\Pi}{dt}=i[{\cal H}'',\bm\Pi]. $$

For particles in a nonstationary electroweak field, it takes the form
\begin{equation}
\begin{array}{c}
\frac{d\bm{\Pi}}{dt} =\left\{\left(\frac{\mu_0m} {\epsilon
'+m}+\mu'\right)\frac{1}{\epsilon '},\left[\bm{\Pi}\times[\bm
E\times\bm \pi]\right]\right\}_++ \left\{\left(
\frac{\mu_0m}{\epsilon '}+\mu'\right),[\bm\Sigma\times\bm
H]\right\}_+\\-  \frac{\mu'}{2}\left\{\frac{1}{\epsilon '(\epsilon
'+m)},\biggl( [\bm\Sigma\times\bm \pi](\bm \pi\cdot\bm
H)+(\bm H\cdot\bm \pi) [\bm\Sigma\times\bm \pi]\biggr)\right\}_+\\
 -\frac{G}{2\sqrt2}\Biggl\{\frac{1}{\epsilon
'},\biggl(C_1\left\{[\bm\Sigma\times\bm \pi],n ({\bm r})\right\}_+
+C_2\left[\bm{\Sigma}\times[\bm{\sigma'}\times\nabla n ({\bm
r})]\right]\biggr)\Biggr\}_+.
\end{array} \label{eq36} \end{equation}

   The corresponding equation for stationary electroweak fields was derived
in \cite{STM}.

The transition to the semiclassical description is also simple.
For free particles, the lower
spinor is equal to zero in the FW representation. For particles in external
fields, the maximum ratio of the lower and upper spinors is of the first
order of $W_{int}/E$, where $W_{int}$ is the energy of the particle
interaction with external fields.
Thus, we obtain $(\chi^{\dag}\chi)/(\phi^{\dag}\phi)\sim(W_{int}/E)^2$.
Therefore, the contribution of the lower spinor is negligible and
the transition to the semiclassical equations is performed by
averaging the operators in the equations for the upper spinor. It is
usually possible to neglect the commutators between the
coordinate and kinetic momentum operators and between different components of
the kinetic momentum operator (see \cite{SJETP2}). As a result, the
operators $\bm\sigma,\bm\sigma'$ and $\bm\pi$ should be substituted by
the corresponding classical quantities: the average spin, $\bm\xi~ (\bm\xi'$
for matter particles), and the kinetic momentum. For the latter quantity
we retain the designation $\bm\pi$. The semiclassical equations of particle
and spin motion are:
\begin{equation}   \begin{array}{c}
\frac{d\bm \pi}{dt}=e\bm E+\frac{e}{\epsilon'}[\bm\pi\times\bm H]-
\frac 12\Biggl(\frac{\mu_0m}{\epsilon'+m}+\mu'\Biggr)\frac{1}{\epsilon'}
\Biggl[2\nabla(\bm\xi\cdot[\bm\pi\times\bm E])-\Delta\bm E\Biggr]\\
-\frac{\mu_0m}{4}\cdot\frac{2\epsilon'^2+2\epsilon' m+m^2}
{\epsilon'^4(\epsilon'+m)^2}(\bm\pi\cdot\nabla)\nabla(\bm
\pi\cdot\bm E)
+\Biggl(\frac{\mu_0m}{\epsilon'}+\mu'\Biggr)\nabla(\bm\xi\cdot\bm H)\\
-\frac{\mu'}{\epsilon'(\epsilon'+m)}\Biggl[(\bm{\xi}
\!\cdot\!\bm\pi)\nabla(\bm{H}\!\cdot\!\bm\pi)+ 2\pi\nabla(\bm
j\!\cdot\! \bm\pi)\Biggr],
\end{array}  \label{eq37} \end{equation}
\begin{equation}   \begin{array}{c}
\frac{d\bm{\xi}}{dt} =2\left(\frac{\mu_0m}
{\epsilon'+m}+\mu'\right)\frac{1}{\epsilon'}\left[\bm{\xi}\times
[\bm{E}\times\bm{\pi}] \right]+2\left(
\frac{\mu_0m}{\epsilon'}+\mu'\right)[\bm{\xi}\times \bm{H}]\\
-\frac{2\mu'}{\epsilon'(\epsilon '+m)}(\bm {H}\cdot\bm
{\pi})[\bm{\xi}\times\bm { \pi}]
-\frac{G}{\sqrt{2}\epsilon'}\biggl(2C_1[\bm{\xi}\times\bm{\pi}]n({\bm
r})+C_2\left[\bm{\xi}\times[\bm{\xi'}\times\nabla   n ({\bm
r})]\right]\biggr).
\end{array}  \label{eq38} \end{equation}

Equation (37) shows that the particle motion depends on the spin orientation. The
corresponding term determines the Stern-Gerlach force.

   It is not as convenient to use the Dirac representation to
derive quantum equations of particle and spin motion in a similar manner.
In this case, it is necessary to
extract the polarization operator, $\bm O$, from the obtained equations.
This problem is rather difficult because the operator $\bm O$ in the Dirac
representation is defined by the cumbersome expression given in \cite{TKR}.

For particles in external fields, the FW transformation also changes the
form of the kinetic momentum operator. In particular, the equation of spin
motion in the Dirac representation depends on the operator $\bm{\pi}_D=
U^{-1} \bm{\pi}U$ just as the corresponding
equation in the FW representation depends on the operator $\bm{\pi}$. However,
these two equations differ in their functional dependence on $\bm{\pi}$.
The use of the FW representation protects from both
an error in derived equations of particle
and spin motion and an incorrect interpretation of these equations.

   Another method of transition to the semiclassical description is based
on the trajectory-coherent solution of the Dirac equation \cite{Ba}.

\section {DISCUSSION}

As mentioned above, the proposed method permits obtaining a transformed
Hamiltonian to
within first-order terms in the field parameters after the first
transformation. In this case, all the other canonical methods need several
transformations \cite{FW,E,KT,GS,Hol,BD,VJ}. Therefore, the method described
above can be successfully used even for solving nonrelativistic problems.
For this purpose, the transformation operator, $U$, can be expanded in a
series of $1/m$. Of course, such an expansion is helpful only in the case of
transformation of the operator ${\cal E}-i(\partial/\partial t)$. The
transformation of other operators leads to the appearance of the term
$\beta\epsilon$ in Eq. (31).

Consider the classical example of the FW transformation for a nonrelativistic
Dirac particle in an electromagnetic field. We calculate the Hamiltonian to
within terms of orders of $(p/m)^4$ and $p^2W/m^3$, where $W$ means
$e\Phi,~e\bm A$ (see
\cite{FW,BD}). In this approximation, the first double commutator in Eq. (29) is
negligible and the second one is equal to
%The formula is corrected
\begin{equation}  \begin{array}{c}
\frac14\left[\frac{\beta{\cal O}}
{\sqrt{2\epsilon(\epsilon+m)}},\left[\frac{\beta{\cal O}}
{\sqrt{2\epsilon(\epsilon+m)}},\left({\cal
E}-i\frac{\partial}{\partial t} \right)\right]\right]\\
=\frac{1}{8m^2}\left[\bm\gamma\cdot\bm\pi,\left[
\bm\gamma\cdot\bm\pi,\left({\cal E}-i\frac{\partial}{\partial t}
\right)\right]\right]=\frac{e}{8m^2}\left[\bm{\Sigma}\cdot(\bm\pi\times\bm
E)-\bm{\Sigma}\cdot(\bm E\times\bm \pi)-\nabla\cdot\bm E\right].
\end{array}  \label{eq39} \end{equation}

As
$$ \epsilon=\sqrt{m^2+(\bm\alpha\cdot\bm\pi)^2}=\sqrt{m^2+\bm\pi^2-e\bm\Sigma
\cdot\bm H}=m+\frac{\bm\pi^2}{2m}-\frac{\bm\pi^4}{8m^3}-\frac{e}{2m}\bm\Sigma
\cdot\bm H, $$
the transformed Hamiltonian is expressed by the well-known formula \cite{FW,BD}:
%The formula is corrected
\begin{equation}   \begin{array}{c}
{\cal
H}''=\beta\left(m+\frac{\bm\pi^2}{2m}-\frac{\bm\pi^4}{8m^3}\right)
+e\Phi-\frac{e}{2m}\bm\Pi\cdot\bm H\\
+\frac{e}{8m^2}\left[\bm{\Sigma}\cdot(\bm\pi\times\bm
E)-\bm{\Sigma}\cdot(\bm E\times\bm \pi)-\nabla\cdot\bm E\right].
\end{array}  \label{eq40} \end{equation}

Thus, the proposed method permits obtaining this formula after the computation
of only one double commutator. All the other canonical methods require cumbersome
calculations \cite{FW,E,Hol,BD,VJ}. For example, the classical method of Foldy and Wouthuysen
require three successive transformations and a calculation of numerous commutators.
The noncanonical methods (the Pauli's elimination method and others)
\cite{AB,STMP,VJ,Pa,BL,BLP,St} permit deriving Hamiltonian (40) in an easier
way. Nevertheless, the proposed method is very simple even compared to them.
Moreover, it gives an opportunity to find a transformed Hamiltonian with any
accuracy even for relativistic particles in external fields.

\section {SUMMARY}

In this work, a method of FW transformation for relativistic particles in
external fields is proposed. This method is simple and reliable. It performs
the exact FW transformation as in known particular cases \cite{E,C,T}, as in
others. This property distinguishes
the proposed method from the other methods developed for relativistic
particles \cite{B,GS,STMP}. The method is based on the well-known
elaborations \cite{FW,BD}.
First, a transformation similar to the FW transformation for free particles
is performed for particles in external fields. Second, a transformation
similar to the FW transformation for nonrelativistic particles is carried out.
In the general case, the FW transformation is approximate.
As an example, the Hamilton operator in the FW representation for
relativistic particles with AMM interacting with nonstationary electroweak fields
is found to within second derivatives of potentials.

%\footnotesize
%\renewcommand{\baselinestretch}{1.4}

\end{document}